\documentclass[%
 aip,
amsmath,amssymb,
 reprint,%
]{revtex4-1}

\usepackage{graphicx}% Include figure files
\usepackage{dcolumn}% Align table columns on decimal point
\usepackage{bm}% bold math
\usepackage[utf8]{inputenc}
\usepackage[T1]{fontenc}
\usepackage{mathptmx}
\usepackage{etoolbox}

\usepackage{siunitx}
\usepackage{graphicx}
\usepackage{dcolumn}
\usepackage{bm}
\usepackage{xcolor}
\usepackage{verbatim}
\usepackage{soul}
\DeclareMathOperator\arctanh{arctanh}

\begin{document}

%\preprint{AIP/123-QED}

\title{Double loop dc-SQUID as a tunable Josephson diode}

\author{A. Greco}
\email{angelo.greco@nano.cnr.it}
\affiliation{NEST, Istituto Nanoscienze-CNR and Scuola Normale Superiore, I-56127 Pisa, Italy}

\author{Q. Pichard}
\affiliation{NEST, Istituto Nanoscienze-CNR and Scuola Normale Superiore, I-56127 Pisa, Italy}

\author{E. Strambini}
\email{elia.strambini@nano.cnr.it}
\affiliation{NEST, Istituto Nanoscienze-CNR and Scuola Normale Superiore, I-56127 Pisa, Italy}

\author{F. Giazotto}
\email{francesco.giazotto@sns.it}
%\homepage{http://web.nano.cnr.it/sqel/}
\affiliation{NEST, Istituto Nanoscienze-CNR and Scuola Normale Superiore, I-56127 Pisa, Italy}

%\date{\today}

\begin{abstract}
The development of superconducting electronics requires careful characterization of the components that make up electronic circuits. Superconducting weak links are the building blocks of most superconducting electronics components and are characterized by highly nonlinear current-to-phase relations (CPR), which are often not perfectly known. Recent research has found that the Josephson diode effect (JDE) can be related to the high harmonic content of the current-to-phase relation of the weak links embedded in superconducting interferometers. This makes the JDE a natural tool for exploring the harmonic content of weak links beyond single-harmonic CPR. 
In this study, we present the theoretical model and experimental characterization of a double-loop superconducting quantum interference device (DL-SQUID) that embeds all-metallic superconductor-normal metal-superconductor junctions. The proposed device exhibits the JDE due to the interference of the supercurrents of three weak links in parallel, and this feature can be adjusted through two magnetic fluxes, which act as experimental knobs. We carry out a theoretical study of the device in terms of the relative weight of the interferometer arms and the experimental characterization concerning flux tunability and temperature.
\end{abstract}

\maketitle
\begin{figure}[t!]
\includegraphics[width=1\columnwidth]{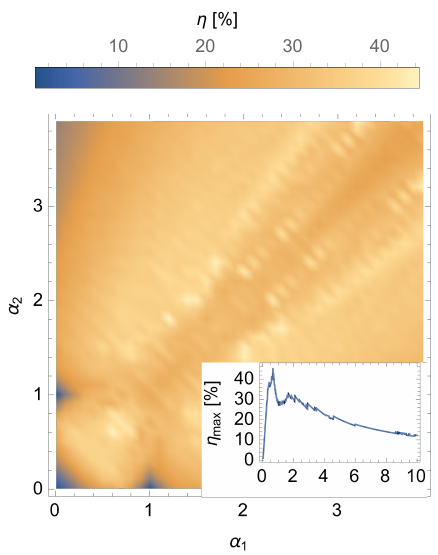}
\caption{Theoretical apical rectification coefficient $\eta_{max}$ as a function of $\alpha_{1,2}$. Inset: theoretical apical rectification coefficient calculated for symmetrical device configurations, hence for $\alpha_1=\alpha_2$.}
\label{fig:max_asymmetries} 
\end{figure}
\begin{figure}[t!]
\includegraphics[scale=0.35]{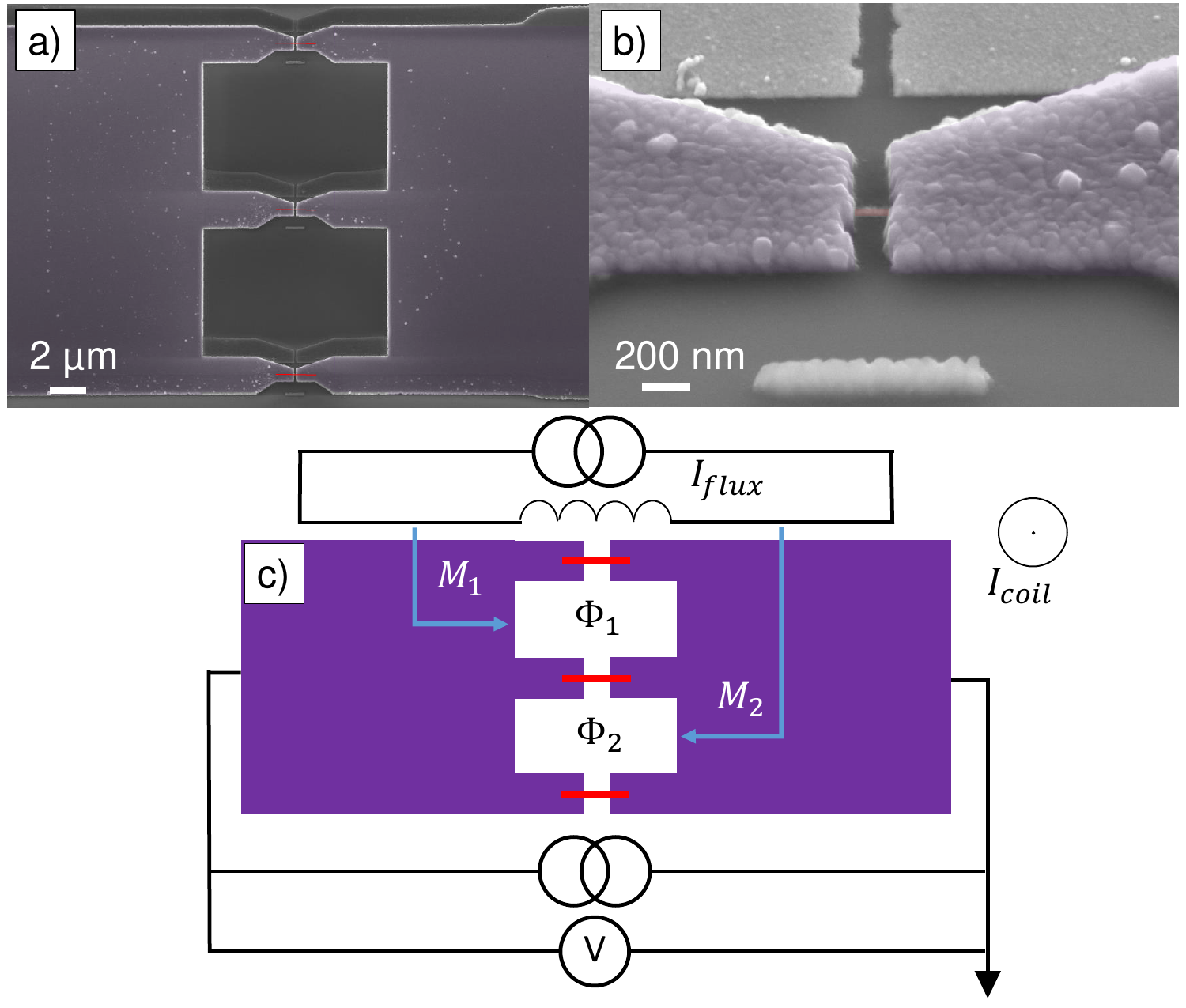}
\caption{a,b) Scanning electron micrographs of a typical real double-loop interferometer. In purple, we highlight the Al layer while in red the Cu nanowires. c) Measurement setup schematics. The purple polygons represent the Al banks and loops of the DL-SQUID, while the red lines are the Cu nanowires. The 4-wire setup for the current vs voltage ($IV$) characterization is indicated by the sketch of the current generator plus the voltmeter. The control parameters for modifying the magnetic fluxes $\Phi_1$ and $\Phi_2$ are the current $I_{coil}$ that generates a homogeneous magnetic field and $I_{flux}$ that generates an inhomogeneous magnetic field through an on-chip superconducting flux line. The mutual inductance between the flux line and the two loops is indicated as $M_1$ and $M_2$.}
\label{fig:device_scheme} 
\end{figure}
\begin{figure}[t!]
\includegraphics[width=1\columnwidth]{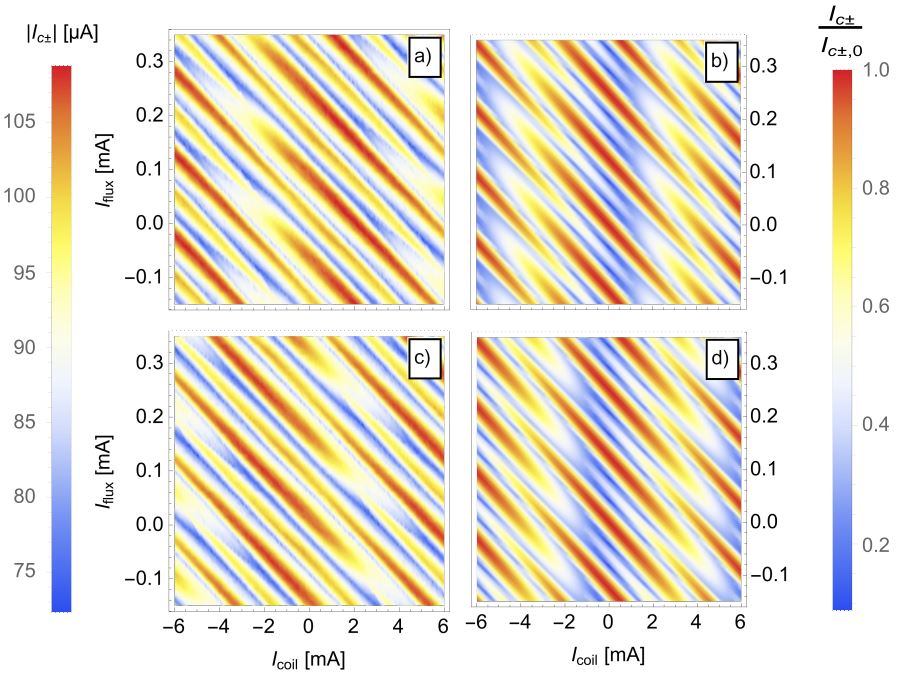}
\caption{Comparison between theoretically calculated and experimentally measured critical currents. Contour plots a) and c) are respectively the measured $I_{c+}$ and $|I_{c-}|$, with their colour bar on the left. Contour plots b) and d) show the theoretically calculated ratio $\frac{I_{c+}}{I_{c+,0}}$ and $\frac{I_{c-}}{I_{c-,0}}$, where $I_{c+,0}$ and $I_{c-,0}$ are the maximum critical currents, with their colour bars on the right.}
\label{fig:experimental_data} 
\end{figure}
\begin{figure}[t!]
\includegraphics[width=1\columnwidth]{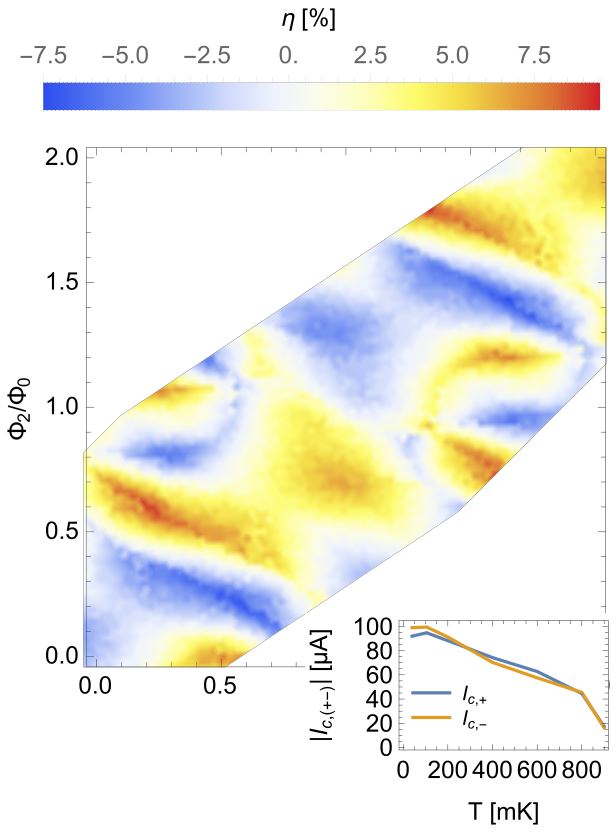}
\caption{Experimental rectification $\eta$ calculated as a function of the reduced fluxes $\Phi_{1}$ and $\Phi_{2}$ in the two loops. The color bar expresses $\eta$ in \%.
Inset: experimental $I_{c,+}$ and $I_{c,-}$ as a function of bath temperature measured for $I_{coil}=$ \SI{200}{\micro\ampere} and $I_{flux}=$ \SI{0}{\micro\ampere}.}
\label{fig:experimental_rectification} 
\end{figure}

In recent years, there has been a growing interest in the development of superconducting electronics \cite{huang2022survey,braginski2019superconductor,de2019josephson,rocci2020gate}. This technology has several interesting applications, such as low energy consumption and extremely high-frequency capabilities \cite{jones2018, chen2022, chen2021, likharev2012, tomita2023, holmes2013}. Additionally, superconducting electronics is a useful platform for studying fundamental physics related to condensed matter \cite{giazotto2006opportunities,likharev1999,fornieri2017towards,paolucci2018phase,pekola2021colloquium,roddaro2011hot}. An example of this is the Supercurrent Diode Effect (SDE), which is the property of superconducting two-terminal devices to transfer dissipationless current with different magnitudes in two directions. This phenomenon originates from a multitude of physical effects, including systems with ferromagnetic layers \cite{strambini2022} and hybrid semiconductor-superconductor junctions\cite{valentini2024,baumgartner2022,mazur2022, coraiola2023, gupta2023}, superconducting quantum interference devices (SQUIDs) \cite{souto2022, souto2023, desimoni2024, ciacca2023, paolucci2023, greco2023}, metallic constrictions \cite{margineda2023_1, margineda2023_2, fominov2022} and metallic strips \cite{satchell2023, hou2023}. It has been demonstrated that in SQUID-like supercurrent diodes, the harmonic content of the weak links forming the interferometer plays a significant role in the appearance of SDE when there is no substantial self-inductance \cite{greco2023, coraiola2023}.
\begin{figure}[t!]
\includegraphics[width=1\columnwidth]{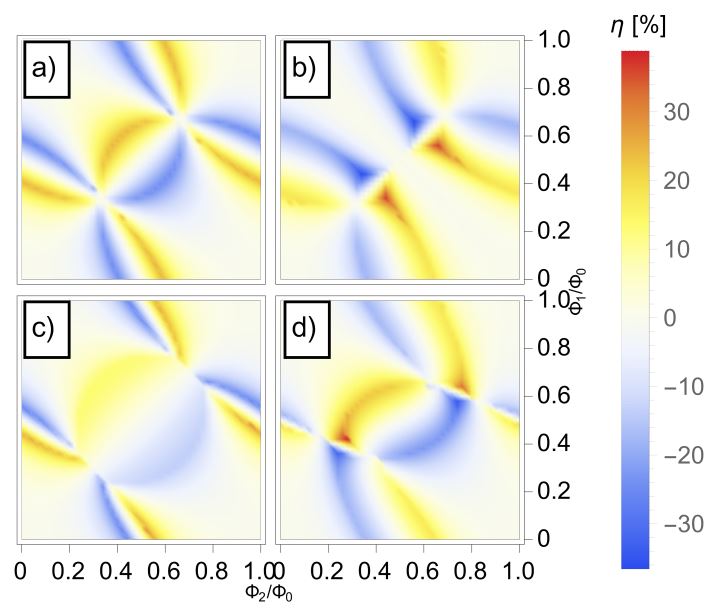}
\caption{Calculated rectification coefficient  $\eta$ as a function of $\Phi_1$ and $\Phi_2$ for different asymmetries of the weak links.
Plot a), $\alpha_1=\alpha_2=1$; plot b), $\alpha_1=\alpha_2=1/2$; plot c), $\alpha_1=\alpha_2=2$; plot d), $\alpha_1=0.8$ and $\alpha_2=1.5$.}
\label{fig:asymmetries} 
\end{figure}
This finding opens up new possibilities for exploring these nonlinearities through SDE, providing a practical tool for investigating the physics of weak links beyond the single-harmonic Current-to-Phase Relation (CPR).\\
In this work, we introduce a non-reciprocal supercurrent element, which is realized through a double-loop SQUID (DL-SQUID), and present its theoretical model and experimental implementation. The device can be controlled by magnetic fluxes that thread a double-loop geometry, allowing interference among its three metallic weak links and tuning of the overall CPR. The double loop structure allows to break the inversion symmetry of the device, enabling non-reciprocal superconducting transport from reciprocal elements. The weak links are realized using superconductor-normal metal-superconductor (SNS) Josephson junctions, which carry a large amount of dissipationless current compared to conventional JJs due to their tunnel-barrier-free structure. We demonstrate theoretically that the presented device can achieve the Josephson Diode Effect (JDE) by breaking both inversion and time-reversal symmetry using external magnetic fields only, producing a theoretical sizable rectification even for a perfectly symmetric device. In our measurement, we can fully explore the parameter space by tuning the magnetic flux of each loop singularly.\\
The DL-SQUID is composed of three weak links arranged in parallel forming two superconducting loops. Each loop is threaded by a magnetic flux, $\Phi_1$ and $\Phi_2$, and the supercurrent flowing through the weak links are indicated as $I_1$, $I_2$ (side junctions), and $I_0$ (central junction). The weak links are supposed to have a CPR that can be described by the Kulik - Omel'yanchuk 1 (KO-1) relation \cite{kulik1975}
\begin{widetext}
\begin{eqnarray}
I_j(\varphi,T)=\frac{\pi\Delta(T)}{eR}\cos(\varphi/2)\int_{\Delta(T)\cos(\varphi/2)}^{\Delta(T)} \frac{\tanh(\epsilon/k_BT)}{\sqrt{\epsilon^2-\Delta^2(T)\cos(\varphi/2)^2}} \,d\epsilon,
\label{eq:KO_T}
\end{eqnarray}
\end{widetext}
where $\Delta(T)$ is the temperature-dependent BCS superconducting energy gap, $T$ is the weak link temperature, $e$ is the electron charge, $R$ is the normal-state resistance of the weak link, $\varphi$ is the phase difference across the weak link, and $k_B$ is the Boltzmann constant. At zero temperature ($T=0$),
Eq. \ref{eq:KO_T} can be simplified as
\begin{equation} I(\varphi)=I_0\cos{\left(\varphi/2\right)}\arctanh{\left[\sin{\left(\varphi/2\right)}\right]},
\end{equation}
where $I_0=\frac{\pi\Delta_0}{eR}$, and  $\Delta_0$ is the superconducting gap at zero temperature.
We can then build the total CPR of the DL-SQUID by summing the contributions of the three parallel branches. Indicating the phase drop across the i-th weak link with $\varphi_i$ and its characteristic current with $I_i=\frac{\pi\Delta_0}{eR_i}$, we can write
\begin{align}
I(\varphi_0,\varphi_1,\varphi_2)=&I_1\cos{\left(\varphi_1/2\right)}\arctanh{\left[\sin{\left(\varphi_1/2\right)}\right]}\nonumber\\
&+I_0\cos{\left(\varphi_0/2\right)}\arctanh{\left[\sin{\left(\varphi_0/2\right)}\right]}\label{eq:approx_CPR}\\
&+I_2\cos{\left(\varphi_2/2\right)}\arctanh{\left[\sin{\left(\varphi_2/2\right)}\right]}\nonumber.
\end{align} 
Given the loop geometry, the phase drops across the weak links are not independent, but follow the fluxoid quantization relations
\begin{align}
  \varphi_1-\varphi_0&=2\pi\frac{\Phi_1}{\Phi_0} \label{eq:phase_1}\\
  \varphi_2-\varphi_0&=-2\pi\frac{\Phi_2}{\Phi_0} \label{eq:phase_2},
\end{align}
which hold in the case of negligible loops self-inductance. By substituting Eq. \ref{eq:phase_1} and \ref{eq:phase_2} into \ref{eq:approx_CPR} we can recast the latter in the functional form $I(\varphi_0,\Phi_1,\Phi_2,\alpha_1,\alpha_2)$, where $\alpha_i=\frac{I_i}{I_0}$, with $i=1,2$, is the dimensionless asymmetry parameter, which takes into account differences among the three critical currents of the branches. Moreover, in order to quantitatively evaluate the non-reciprocal transport properties, we define the rectification coefficient $\eta=\frac{I_{c+}-I_{c-}}{I_{c+}+I_{c-}}$, where $I_{c+}=Max_{\varphi0}[I]$ and $I_{c-}=Min_{\varphi0}[I]$ are the absolute maxima and minima of the CPR.\\
Fig. \ref{fig:asymmetries}, displays $\eta(\Phi_1,\Phi_1)$ for four different values of $\alpha_{1,2}$. The reported asymmetry values are as follows: $\alpha_1=\alpha_2=1$ (panel a), $\alpha_1=\alpha_2=1/2$ (panel b), $\alpha_1=\alpha_2=2$ (panel c), and $\alpha_1=0.8$ and $\alpha_2=1.5$ (panel d). This selection of values aims to represent some physical cases that can occur in a real device. For instance, panel a) illustrates the rectification efficiency of a fully symmetric device, which means a device with three identical junctions. Notably, rectification is present even in a perfectly symmetrical device, with an apical value of approximately 25\%. This is a remarkable fact since to achieve JDE or SDE, the inversion symmetry has to be broken, usually by introducing geometric or intrinsic asymmetries in the system. In the presented interferometer, the inversion symmetry can be broken, even for symmetrical weak links configurations, just through a non-uniform magnetic field by making $\Phi_1$ and $\Phi_2$ unequal. Furthermore, we notice two main symmetries of the system that can be found in the present plot a). The first one is an odd symmetry with respect to the diagonal $\Phi_1=\Phi_2$. This means that the transformation $\Phi_1 \rightarrow \Phi_2$ and $\Phi_2 \rightarrow \Phi_1$
produces a CPR which is odd with respect to the initial one, hence $I(\varphi_0,\Phi_1,\Phi_2)=-I(-\varphi_0,\Phi_2,\Phi_1)$. The odd rectification pattern $\eta(\Phi_1,\Phi_2)=-\eta(\Phi_2,\Phi_1)$ is a direct consequence of this fact. The second appreciable symmetry is an even symmetry with respect to the line $\Phi_1=-\Phi_2$. Now, the transformation
$\Phi_1 \rightarrow -\Phi_2$ $\Phi_2 \rightarrow -\Phi_1$
produces an even CPR with respect to the initial one, which in this case leads to $I(\varphi_0,\Phi_1,\Phi_2)=I(\varphi_0,-\Phi_2,-\Phi_1)$. The even symmetry rectification $\eta(\Phi_1,\Phi_2)=\eta(-\Phi_2,-\Phi_1)$ descends from this property. We observe a similar pattern in plots b) and c), where the lateral junctions are equal but have lower and higher critical current values respectively when compared to the central junction. In both cases, we see the same symmetries as in plot a), but with different apical rectification values. Plot b) shows a maximum achievable rectification of 36\%, while plot c) shows 24\%. It's worth noting that the plots shown so far have been for the symmetric device case, where both time-reversal and inversion symmetries are achieved through specific configurations of the magnetic field. However, the last plot on the panel shows a case where all the junctions have different critical currents. This results in the loss of the above-mentioned symmetries but is closer to the physical case. 
It's important to note that, due to the geometrical asymmetry of the junctions, diode effects are visible even at $\Phi_1=\Phi_2$. In the following section, we investigate the impact of these asymmetries on device performance by evaluating the apical rectification $\eta_{Max}= Max_{\phi_1,\phi_2}[\eta]$ as a function of $\alpha_1$ and $\alpha_2$.\\
The contour plot in Fig. \ref{fig:max_asymmetries} illustrates the apical rectification value $\eta_{max}$ for a particular $\alpha_1$ and $\alpha_2$ configuration in the range $0<\alpha_i<4$. $\eta_{max}$ is calculated for each configuration of $\alpha_1$ and $\alpha_2$ by sampling $\eta(\Phi_1,\Phi_2)$ with $6400$ points and by extracting the maximum value in the calculated table.
From the plot we notice that the system reaches rectification as high as $\sim 42\%$ for particular asymmetry values, while for some $\alpha_{1,2}$ values we obtain some areas vanishing rectification. We can see one for $\alpha_1=\alpha_2=0$, which corresponds to the case where the lateral junctions have vanishing critical currents, representing the case of a single weak link. Other two vanishing rectification points can be found for $\alpha_1=0$ and $\alpha_2=1$ and vice versa, which corresponds to the case where one lateral junction has a vanishing critical current while the other is identical to the central one. The latter case corresponds to the single-loop geometry treated in \cite{greco2023}, where we show that with identical branches  $\eta=0$.\\
As already anticipated, the DL-SQUID shows the intriguing feature of being able to break both inversion and time-reversal symmetry using magnetic field biasing only. This feature is achieved due to the presence of the double-loop geometry, that even in a perfectly symmetric weak-link configuration, can induce inversion symmetry breaking if $\Phi_1\neq\Phi_2$. It is hence interesting to investigate $\eta_{max}$ on the symmetry line $\alpha_1=\alpha_2=1$, hence where the side weak links are equal and the central one can vary freely. This is displayed in the inset of Fig. \ref{fig:max_asymmetries}. Here the blue curve shows the raw numerical sampling operated on each weak link configuration with $10000$ points on the $\eta(\Phi_1,\Phi_2)$ function. The black dashed curve is then a moving average on $3$ points superimposed on the raw data, to smooth out the sharp features coming from the raw numerical sampling. $\eta_{max}$ vanishes at $\alpha_{1,2}=0$, corresponding to a case where the side weak links are negligibly small with respect to the central one, then has an absolute maximum at $\alpha_{1,2}\approx0.65$ and finally a second peak at $\alpha_{1,2}\approx1.65$, then going monotonically to zero for higher asymmetry values corresponding to the limit of a single symmetric SQUID.\\
From this analysis, we can deduce some useful concepts. The JDE in this system stems from the high harmonic content of the CPRs used to model the weak links. As such, the harmonic interference of the three branches, together with their relative weights, determines the magnitude of $\eta_{max}$ and the respective $\alpha_{1,2}$ at which the effect appears. Indeed, it can be shown that in the same system, by replacing a KO-1 CPR with a sinusoidal CPR, the JDE is lost regardless of the critical current relative values. By comparing the highest rectification value theoretically obtained in this work, $\eta_{max}=42\%$, with the single-loop version investigated in \cite{greco2023}, which reaches $\eta_{max}=26\%$, we conclude that the JDE here achieved greatly benefits from the addition of another weak link. This can be understood by a much more efficient harmonic combination arising from the presence of the third weak link.\\
Motivated by these theoretical results we fabricated and tested a physical implementation of the above-described DL-SQUID. The device was realized using the standard Niemeyer-Dolan technique, using a double-layer lithographic mask of MMA + PMMA A4. The deposition of the metal layers was performed using a UHV e-beam evaporator with a residual base pressure of $4.6\times 10^{-10}$ Torr. During the first evaporation, we deposited a thin Cu layer of $24$ \si{\nano\metre} at an angle of $35\si{\degree}$, with a rate of $2$\si{\angstrom/\second} and deposition pressure $<10^{-11}$ Torr. Afterward, a second deposition of $180$ \si{\nano\metre} of Al at $0 \si{\degree}$ and deposition pressure of $10^{-8}$ Torr was performed to realize the thick Al banks and loops. The Al deposition rate was kept at $1$ \si{\angstrom/\second} for the first $30$ \si{\nano\metre} and then raised to $2$ \si{\angstrom/\second} until the end of the evaporation.\\
Two micrographs of the DL-SQUID are shown in Fig. \ref{fig:device_scheme} a) and b). In image a), the Al loops are highlighted in purple, while the Cu nanowires are indicated in red. Image b) is a zoomed-in view of the Cu nanowire embedded into the Al banks. This image shows the high aspect ratio of the structure used to create the three-dimensional (3D) constrictions. The weak links in this device have a length between \SI{90}{\nano\metre} and \SI{110}{\nano\metre}.
Figure \ref{fig:device_scheme} c) illustrates a sketch of the measurement setup used to control the magnetic fluxes piercing the loops and to measure the switching currents as a function of the latter. 
The fluxes $\Phi_1$ and $\Phi_2$ are controlled through two different magnetic fields generated by a superconducting coil, placed at the bottom of the dilution refrigerator, and an on-chip superconducting flux line. The current passing through the coil and the flux line are indicated respectively as $I_{coil}$ and $I_{flux}$ and are our control parameters. The relation between $\Phi_{1,2}$ and the control parameters can be expressed through mutual inductances by the linear system
\begin{align}
    \Phi_1=&M_1 I_{flux} + M_{coil} I_{coil} \label{eq:flux_1}\\
    \Phi_2=&M_2 I_{flux} + M_{coil} I_{coil} \label{eq:flux_2}
\end{align}
where $M_1$ and $M_2$ are the mutual inductances between the flux line and the loop embracing $\Phi_1$ and $\Phi_2$, respectively. 
%where the mutual inductances $M_1$ and $M_2$ are respectively the ones between the flux line and the loop related to $\Phi_1$ and $\Phi_2$. 
$M_{coil}$ is the mutual inductance between the superconducting coil and the two loops and can be considered equal due to the uniformity of the field and the equal size of the two loops. 
%In this linear system, the mutual inductances between the superconducting coil and the two loops are considered equal. 
Figure \ref{fig:device_scheme} c) also shows a current generator and a voltmeter, used for the 4-wire measurements of the switching currents. By using the measurement scheme just described one can measure the critical currents of the DL-SQUID as a function of the control parameters, and in this way test our theoretical predictions.\\
Figure \ref{fig:experimental_data} displays the comparison between the measured and theoretically calculated values of the critical currents in our DL-SQUID. The panel shows on the left the experimental data representing $I_{c+}$ a), $I_{c-}$ c) and their relative colour bar indicating the critical current in \si{\micro\ampere}.
On the right panels the theoretically calculated values of $\frac{I_{c+}}{I_{c+,0}}$ b) and $\frac{I_{c-}}{I_{c-,0}}$ d) are reported for $\alpha_1=\alpha_2=1$. There, $I_{c+,0}$ and $I_{c-,0}$ are the absolute maximum and minimum critical currents evaluated in the parameter space, respectively. 
To reproduce the experimental data shown in a) and c) we made use of Eqs. (\ref{eq:flux_1}) and (\ref{eq:flux_2}) to link $\Phi_1$ and $\Phi_2$ with the experimental control parameters $I_{coil}$ and $I_{flux}$. The mutual inductance between the coil and the loops turns out to be $M_{coil}=$ \SI{16}{\pico\henry}, while the mutual inductances between the flux line and the two loops result in $M_{1}=$ \SI{0.9}{\pico\henry} and $M_{2}=$ \SI{0.6}{\pico\henry}, respectively. 
The latter were estimated from the analytical formula of the mutual inductance between a strip and a square spiral, while the former from the periodicity of the interference pattern of the device. By examining the theoretical and experimental results we can notice some aspects that deserve deeper consideration. First of all, the predictions for $\frac{I_{c+}}{I_{c+,0}}$ and $\frac{I_{c-}}{I_{c-,0}}$ reveal a theoretical maximum suppression of the critical currents of about $\frac{I_{c+-}}{I_{c+-,0}}\approx90\%$. This is not the case for the experimental data, where we can record a maximum suppression of about $35\%$. The reason for this inefficiency in suppressing the critical current may be related to the length of the weak links, which for the present device spans from \SI{90}{\nano\metre} to \SI{110}{\nano\metre}. The length of the weak link ($L$) for a SNS junction indeed determines the harmonic content of its CPR \cite{heikkila2002}, and depending on the ratio $L/\xi_0$, where $\xi_0=\sqrt{\hbar D/\Delta_{w,0}}$ is the coherence length in the Cu nanowires, one can sit in the multi-harmonic or the sinusoidal CPR regime. For Cu proximitized by Al banks, we set $D\approx$ \SI{80}{\centi\metre^2/\second} as the diffusion coefficient of Cu, and $\Delta_{w,0}=1.764 k_B T_c\approx$ \SI{200}{\micro\electronvolt} as the superconducting gap of Al, leading to a coherence length of $\xi_0\sim$ \SI{50}{\nano\metre}. As a consequence, the ratio $L/\xi_0$ spans from $\approx1.8$ to $\approx2.2$ thereby setting the frame of the intermediate-length junction regime. Our theoretical model based on the KO-1 relation (see Eq. \ref{eq:KO_T}) is valid strictly in the \textit{short junction} limit, i.e., $L/\xi_0\ll 1$, and therefore does not take into account the variation of the harmonic content which happens when the junction is longer. A reduction in the harmonic content of the weak links, which takes place in the \textit{long} regime, reflects in a reduced interference among the branches of the DL-SQUID. This said, and considering the unknown asymmetries among the branches, we can appreciate that the experimental data are qualitatively reproduced by our theoretical model if one considers the overall behaviour of the curves.\\
The experimental rectification $\eta(\Phi_1, \Phi_2)$ as a function of the fluxes in the two loops is shown in Fig. \ref{fig:experimental_rectification}.
The contributions of the single fluxes were de-embedded from the experimental parameters by inverting Eqs. \ref{eq:flux_1} and \ref{eq:flux_2}. This allows us to compare the experiment with the simulated value reported in Fig.\ref{fig:asymmetries}. The data extend mainly around the $\Phi_1=\Phi_2$ line since this is in the parameter region explored by sweeping the current in the superconducting coil. Here, $\eta$ ranges from $+ 8\%$ to $- 8\%$ in the explored fluxes with a clear periodic feature in the flux quantum. Its behaviour is not compatible with the symmetric configurations reported in Fig.\ref{fig:asymmetries} suggesting an asymmetry between the three junctions, as reported in Fig.\ref{fig:asymmetries}d).\\
As already mentioned, the reduced rectification can be explained by assuming that the harmonic content of our weak links is reduced with respect to the short limit expressed by the KO-1 model. In this sense,  the harmonic content results to be a key ingredient for JDE. This concept can be well understood by studying the temperature behaviour of $I_{c,+}$ and $I_{c,-}$, since these two quantities are directly related to the harmonic mixing among the weak links of the DL-SQUID, and from Eq.\ref{eq:KO_T} we know that the harmonic content in the CPR is directly related to the bath temperature. The inset of Figure \ref{fig:experimental_rectification} shows $I_{c,+}$ and $I_{c,-}$ as a function of bath temperature measured between \SI{30}{\milli\kelvin} and \SI{900}{\milli\kelvin} for the $\Phi_{1,2}$ configuration given by $I_{coil}=$ \SI{200}{\micro\ampere} and $I_{flux}=$ \SI{0}{\micro\ampere}. In this plot we notice an interesting feature, that is a magnitude inversion between $I_{c,+}$ and $I_{c,-}$ as a function of temperature. This feature can be observed at \SI{270}{\milli\kelvin} and at \SI{800}{\milli\kelvin}, and this fact is accompanied by a sign change of $\eta$. This behaviour can be explained by considering that the three weak links have different lengths and as a consequence different weights in their CPRs. It can be shown with our theoretical model that a DL-SQUID, for certain $\Phi_{1,2}$ configurations, shows sign changes of $\eta$ as a function of temperature. This characteristic feature is present when at least three weak links interfere, indeed it was not observed in the single-loop version studied in \cite{greco2023}.\\
In conclusion, we have demonstrated a double-loop interferometer based on Al/Cu/Al superconductor-normal metal-superconductor weak links. The KO-1 formula describes these weak links, which provide a multi-harmonic CPR  yielding sizable JDE when inserted into a SQUID-like structure. In our research, we used rectification $\eta$ as a tool to examine the anharmonicity of the weak links within the interferometer. We explored the tunability of the device with respect to magnetic fluxes in the double-loop geometry, as well as its temperature behaviour. Our findings indicate that the device exhibits supercurrent rectification sign changes with temperature, which is a signature of the junctions harmonic content tunability. We hope that this work will encourage the use of JDE as a way to investigate nonlinear superconducting building blocks, leading to further developments in superconducting electronics.\\

The data that support the findings of this study are available from the corresponding author upon reasonable request.

\begin{acknowledgments}
We acknowledge the EU’s Horizon 2020 Research and Innovation Framework Programme under Grant No. 964398 (SUPERGATE), No. 101057977 (SPECTRUM), and the PNRR MUR project PE0000023-NQSTI for partial financial support.
\end{acknowledgments}

\nocite{*}
\bibliography{aipsamp}% Produces the bibliography via BibTeX.

\end{document}